\documentstyle [12pt,dina4,psfig]{article}
\begin{document}
\parindent=0cm
\parskip=1.5mm

\def\bi{\begin{list}{$\bullet$}{\parsep=0.5\baselineskip
\topsep=\parsep \itemsep=0pt}}
\def\ei{\end{list}}

\def\phi{\varphi}
\def\-{{\bf --}}
\def\vm{v_{max}}
\def\ta{\tilde a}
\def\ts{\tilde s}
\def\tr{\tilde r}
\newcommand{\s}{\sigma}
\newcommand{\la}{\lambda}
\newcommand{\eps}{\varepsilon}
\newcommand{\al}{\alpha}
\newcommand{\nn}{{\cal N}}

\newcommand{\be}{\begin{equation}}
\newcommand{\ee}{\end{equation}}
\newcommand{\bea}{\begin{eqnarray}}
\newcommand{\eea}{\end{eqnarray}}

\newcommand{\nonu}{\nonumber\\}

\renewcommand{\thefootnote}{\fnsymbol{footnote}}

\parindent=0cm
\begin{center}
  {\LARGE\bf Garden of Eden states in traffic models}
\end{center}
\vskip1.8cm \renewcommand{\thefootnote}{\fnsymbol{footnote}}
\setcounter{footnote}{1}
\begin{center}
  {\Large Andreas Schadschneider$^{1}$ and Michael Schreckenberg$^{2}$}
\end{center}
\vskip1.3cm
\begin{center}
  $^{1}$ Institut f\"ur Theoretische Physik\\ Universit\"at zu K\"oln\\ 
  D--50937 K\"oln, Germany\\
  email: {\tt as@thp.uni-koeln.de}
\end{center}
\begin{center}
  $^{2}$ Theoretische Physik/FB 10\\ 
  Gerhard-Mercator-Universit\"at Duisburg\\ 
  D--47048 Duisburg, Germany\\
  email: {\tt schreck@uni-duisburg.de} 
\end{center}
\begin{center}
\today
\end{center}

\vskip1.3cm \vskip2cm {\large \bf Abstract}\\[0.2cm]
We investigate the allowed configurations in the stationary state of
the cellular automaton model for single-lane traffic. It is found that
certain states in the configuration space can not be reached if one uses
parallel dynamics. These so-called {\em Garden of Eden}
(GoE) states do not exist for random-sequential dynamics and are 
responsible for the strong short-ranged correlations found in parallel
dynamics. By eliminating the GoE states we obtain a simple and effective
approximative description of the model. For $\vm=1$ the exact solution is
recovered. For $\vm=2$ this elimination leads to much higher
values of the flux compared to the mean-field result which are in good
agreement with Monte Carlo simulations.
\hspace{.4cm} 

\vskip2cm
\vfill
\pagebreak

\section{Introduction}

The description of traffic flow using cellular automata (CA) is quite
successful \cite{Jue}, despite the simplicity of the model \cite{ns}. CA are,
by design, ideal for large-scale computer simulations. This fact has already
been used for the simulation of urban traffic in various cities,
see e.g.\ \cite{duis,dallas1,dallas2}. On the other hand, analytical
descriptions are difficult. In \cite{ss,ssni,comf} we have developed
several methods which yield an approximate description of the stationary
state. In certain limits (e.g.\ $\vm=1$ or $p\to 0$), these methods even 
become exact (see \cite{as} for a review). These approaches are based on 
a microscopic description which takes into account certain correlations.
Part of the difficulties come from the fact that one uses parallel dynamics.
This introduces a non-local aspect into the problem since the whole
lattice is updated at once. On the other hand, random-sequential dynamics
is much simpler to treat analytically. For $\vm=1$ for instance, simple
mean-field theory gives already the correct steady state, i.e.\ there
are no correlations. Here we propose a rather simple analytical approach
which exhibits the main difference between parallel and random-sequential
dynamics very clearly. For parallel dynamics not all states of the
configuration space can be reached by the dynamics, some are 'dynamically
forbidden'. This is not the case for random-sequential dynamics.

For completeness we recall the definition of the CA model for
single-lane traffic \cite{ns}. The street is divided 
into $L$ cells which can be occupied by at most one car or be empty. 
The state of each car is described by an internal parameter ('velocity') 
which can take on only integer values $v=0,1,2,\ldots,\vm$. 
The state of the system at time $t+1$ can be obtained from the
state at the previous time $t$ by applying the following four simple
update steps to all cars at the same time (parallel dynamics):
\begin{description}
\item[R1] Acceleration:\ \ \  $v_j(t)\rightarrow v_j(t+\frac{1}{3})
=\min\{v_j(t)+1,\vm\}$
\item[R2] Braking:\ \ \  if\ \  $v_j(t+\frac{1}{3}) > d_j(t)$\ \ \ then\ \ \  
$v_j(t+\frac{1}{3}) \rightarrow v_j(t+\frac{2}{3})=d_j(t)$ 
\item[R3] Randomization:\ \ \  $v_j(t+\frac{2}{3})\ 
{\stackrel{p}{\rightarrow}}\ v_j(t+1)=\max\{0,v_j(t+\frac{2}{3})-1\}$\\
\phantom{ Randomization:} with probability $p$ 
\item[R4] Driving:\ \ \  car $j$ moves $v_j(t+1)$ cells.
\end{description}
Here $d_j(t)$ denotes the number of empty cells in front of car $j$, i.e.\ 
the gap or headway. In the following it will be important that $v_j(t)$
is just the number of cells that car $j$ moved in the timestep $t-1\to t$.


In \cite{ssni} we investigated a microscopic mean-field theory (MFT) for 
this kind of cellular automaton. The most important result was that the 
flows obtained are much too small compared to computer
simulations. The reason is that MFT cannot account for
the 'particle-hole attraction' found in the stationary state, i.e.\
the probability to find an empty cell in front of a (moving) car
is {\em enhanced} compared to the mean-field result.
This effect is taken into account by the $n-$cluster approximation introduced
in \cite{ss,ssni}. Here the lattice is divided into clusters of length $n$
which overlap $n-1$ cells\footnote{MFT correponds to $n=1$.}. It turned out 
that the 2-cluster approximation is {\em exact} for $\vm=1$. For $\vm=2$ 
already small cluster sizes (e.g.\ $n=4$) gave an excellent agreement 
with numerical results for the flow.

In \cite{comf} an alternative analytic approach was introduced, the
so-called car-oriented mean-field theory (COMF). Here the dynamical
variables are not the occupancies of the cells but the gaps $d_j(t)$ between
consecutive cars. In the COMF these gaps are treated as being independent.
Again we found that COMF is exact for $\vm=1$. For $\vm=2$ and the
'traditional' value $p=0.5$ the flow obtained from COMF is comparable
to the 3-cluster result. However, the COMF seems to become exact in
the limit $p\to 0$, in contrast to the 2-cluster approach \cite{as}.

\begin{figure}[ht]
\centerline{\psfig{figure=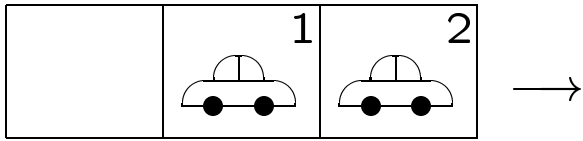,bbllx=150pt,bblly=435pt,bburx=460pt,bbury=520pt,height=2.5cm}}
\caption{A Garden of Eden state for the model with $\vm \geq 2$.}
\label{fig_GoEconf}
\end{figure}

In the present paper we like to present a rather simple extension of
MFT. The key idea is a reduction of the configuration space by
removing all states which cannot by reached dynamically. In the context
of cellular automata these states are called {\em Garden of Eden (GoE) states}
or {\em paradisical states}. A simple example (see Fig.\ \ref{fig_GoEconf})
for $\vm =2$ is the configuration $(\bullet,1,2)$ of two consecutive cells, 
where '$\bullet$' denotes an empty cell and the numbers correspond to the 
velocities of the cars. Cars move from left to right. Obviously the velocity 
is just the number of cells the car moved in the previous timestep. 
Therefore the configuration $(\bullet,1,2)$ could have evolved only
from a state which has two cars in the leftmost cell. Since 
double-occupations are not allowed in the present model, states containing
$(\bullet,1,2)$ are dynamically forbidden, i.e.\ are GoE states.

Here we reinvestigate MFT for $\vm=1$ and $\vm=2$, but eliminate the
GoE states. This {\em paradisical mean-field (pMF)} theory will lead to 
a considerable improvement of the results.

\section{Mean-field theory}

Here we briefly review the MFT results for $\vm=1$ and $\vm=2$. A more 
complete account along with a detailed derivation can be found in 
\cite{ssni}. The densities of cars with velocity $v$ ($v=0,1,\ldots,\vm$) 
in the stationary state is denoted by $c_v$. Therefore the full density of 
cars is given by 
$c=\sum_{v=0}^{\vm} c_v$. For convenience we also introduce the 'hole' 
density $d=1-c$ and the abbreviation $q=1-p$. The flux (or current) is 
given by $f(c)=\sum_{v=1}^{\vm} vc_v$.

For $\vm=1$ the mean-field equations then read:
\bea
c_0 &=& (c+pd)c,\label{mf1a}\\
c_1 &=& qcd \label{mf1b}
\eea
so that the flux is simply given by 
\be
f_{MF}^{(1)}(c)=c_1=qc(1-c).
\ee

For $\vm=2$ the rate equations for the densities are given by
\bea
c_0 &=& (c+pd)c_0 + (1+pd)c(c_1+c_2),\label{mf2a}\\
c_1 &=& d\left[qc_0 + (qc+pd)(c_1+c_2)\right],\label{mf2b}\\
c_2 &=& qd^2(c_1+c_2).\label{mf2c}
\eea
The solution
\bea
c_0 &=&\frac{(1+pd)c^2}{1-pd^2},\\
c_1 &=&\frac{q(1-qd^2)dc}{1-pd^2},\\
c_2 &=&\frac{q^2d^3c}{1-pd^2},
\eea
yields for the flux
\be
f_{MF}^{(2)}(c)=c_1+2c_2=\frac{q(1+qd^2)dc}{1-pd^2}.
\ee

Comparison with the results from Monte Carlo simulations shows that the
MFT results underestimate the flow considerably \cite{ssni}. Therefore 
strong short-range correlations exist which increase the flow
compared to the prediction of MFT. Using the exact solution for the case
$\vm=1$ \cite{ss} one can demonstrate explicitely the existance of a strong 
particle-hole attraction, i.e.\ the probability to find an empty
cell in front of an occupied cell is enhanced compared to the MFT result.

At first sight, this result is surprising. For random-sequential 
dynamics\footnote{In random-sequential dynamics in each timestep a cell 
which is updated is picked at random.} it is known \cite{ns} that
MFT is exact for $\vm=1$. Therefore the origin of the correlations is the 
parallel update procedure. In the following we will see that the existence of
the GoE states is responsible for the differences between parallel and
random-sequential dynamics.


\section{GoE states for $\vm =1$}

The question, whether a state is a GoE state or not, can be decided
locally by investigating just nearest-neighbour configurations.
It turns out that GoE states are all states containing the 
local configurations $(0,1)$ or $(1,1)$, i.e.\ a moving vehicle is
directly followed by another car. This is not possible as can be seen
by looking at the possible configurations at the previous timestep.
The momentary velocity gives the number of cells that the car moved
in the previous timestep. In both configurations the first car
moved one cell. Therefore it is immediately clear that $(0,1)$ is a
GoE state since otherwise there would have been a doubly-occupied
cell before the last timestep. The configuration $(1,1)$ is also
not possible since both cars must have occupied neighbouring cells
before the last timestep too. Therefore, according to rule R2, the
second car could not move.

We now modify the MFT equations (\ref{mf1a}) and (\ref{mf1b}) to take
into account the existence of GoE states. Following the procedure described
in \cite{ssni} it turns out that only
(\ref{mf1a}) has to be modified. Due to this modification the
normalization $c_0+c_1=c$ is no longer satisfied automatically.
Therefore a normalization constant $\nn$ has to be introduced.
The final equations are than given by
\bea
c_0&=&\nn (c_0+pd)c ,\label{pmf10}\\
c_1&=&\nn qcd ,\label{pmf11}
\eea
with the normalization
\be
\nn =\frac{1}{c_0+d}.
\ee
Since $c_0+c_1=c$ we have only one independent variable for fixed density
$c$, e.g.\ $c_1$. Solving (\ref{pmf10}), (\ref{pmf11}) for $c_1$ we obtain
\be
c_1 = \frac{1}{2}\left(1-\sqrt{1-4q(1-c)c}\right).
\ee
The flow is given by $f(c)=c_1$ and we recover the exact solution 
for the case $\vm=1$ \cite{ss}.

This result confirms the expectations mentioned above. One can see
clearly that the difference between random-sequential and parallel
dynamics is the existence of GoE states in the latter. After eliminating
these GoE states, no correlations are left in the reduced configuration 
space.

\section{GoE states for $\vm =2$}

In this case more GoE states exist. In order to identify the GoE states
it is helpful to note that the rules R1--R4 imply $d_j(t)=d_{j}(t-1)
+v_{j+1}(t)-v_j(t)$ and therefore
\bea
d_j(t)&\geq& v_{j+1}(t)-v_j(t),\label{distcond1}\\
v_j(t)&\leq& d_{j}(t-1)\label{distcond2}
\eea
The second inequality (\ref{distcond2}) is a consequence of R2.

In the following we list the elementary GoE states, i.e.\ the 
local configurations which are dynamically 
forbidden (cars move from left to right): 
\bea
(0,1),\quad (0,2),\quad (1,2),\quad (0,\bullet,2),
\label{goe1}\\
(1,1),\quad (2,1),\quad (2,2),\quad (1,\bullet,2),\quad (2,\bullet,2),
\label{goe2}\\
(0,\bullet,\bullet,2).\label{goe3}
\eea
The numbers give the velocity of a vehicle in an occupied cell and
$\bullet$ denotes an empty cell. Using Monte Carlo simulations
we have checked that there are no further elementary GoE states for
clusters up to 10 cells.

The elementary GoE states in (\ref{goe1}) violate the inequality 
(\ref{distcond1}), and the configurations in (\ref{goe2}) violate 
(\ref{distcond2}). The state in (\ref{goe3}) is a second order GoE state. 
Going one step back in time leads to a first order GoE state since
$(0,\bullet,\bullet,2)$ must have evolved from $(0,v)$ (with $v=1$ or $v=2$).

Again we can derive the pMF equations by modifying the method 
for the derivation of the MF mean-field theory \cite{ssni}.
Taking into account only the first order GoE states (\ref{goe1}) and 
(\ref{goe2}) one obtains the following pMF equations:
\bea
c_0 &=& \nn \left[ c_0c+pd(c_0+c_1c)\right],\label{c0eq}\\
c_1 &=& \nn \left[ pd^2(c_1+c_2)+qd(c_0+c_1c)\right],\label{c1eq}\\
c_2 &=& \nn qd^2(c_1+c_2)\label{c2eq}.
\eea
The normalization $\nn$ ensures $c_0+c_1+c_2=c$ and is given
explicitly by
\be
\nn = \frac{1}{c_0+dc_1+d^2c_2}=\frac{1}{c_0+d(1-c_2)}.
\ee

Using (\ref{c2eq}) we can express $c_2$ through $c_0$ and $c$ only:
\be
c_2 = \frac{1}{2d}\left(c_0+d-\sqrt{(c_0+d)^2-4qd^3(c-c_0)}\right).
\label{c2}
\ee
Inserting this result into (\ref{c0eq}) we obtain a cubic equation
which determines $c_0$ in terms of the parameters $c$ and
$p$:
\be
\alpha c_0^3 +\beta c_0^2 + \gamma c_0 + \delta = 0
\label{poly}
\ee
where the coefficients are given by
\bea
\alpha &=& -qd^3-pd^2+qd-q,\\
\beta &=& (3p^2-p-1)d^4+(-p^2-3p+1)d^3+(-p^2+p+2)d^2-3qd+q,\\
\gamma &=& pcd[(-p^2-3p+2)d^3+(p^2+p+1)d^2+(2p-3)d+q],\\
\delta &=&-(pcd)^3 .
\eea
In principle the zeroes of (\ref{poly}) can be determined explicitely.
The physical solution is the one in the interval $[0,1]$. 
The flow can be calculated as $f(c,p)=c_1+2c_1$ where $c_1$ and $c_2$
are determined by (\ref{c2}) and $c=c_0+c_1+c_2$.

\begin{figure}[ht]
\centerline{\psfig{figure=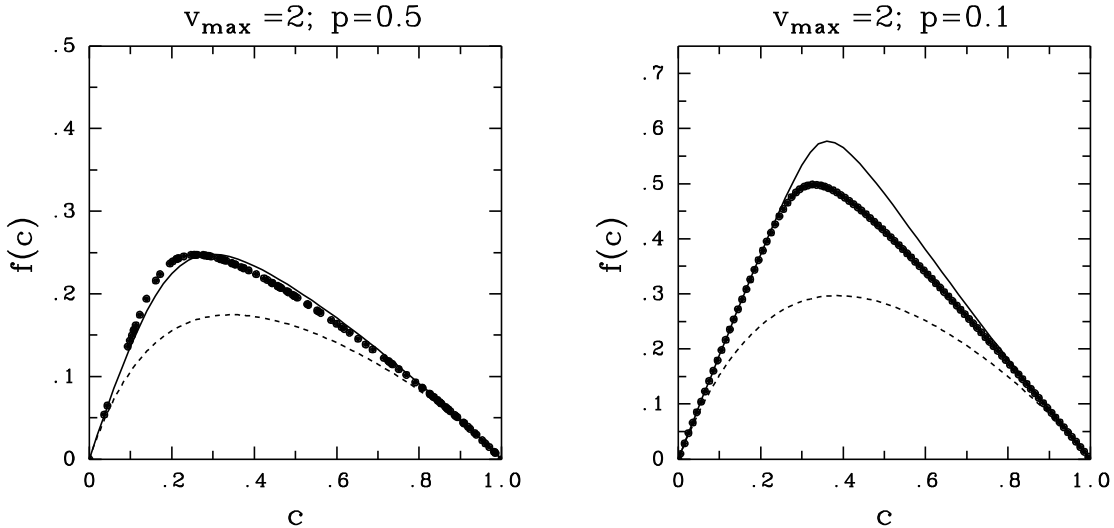,bbllx=35pt,bblly=420pt,bburx=580pt,bbury=685pt,height=7.5cm}}
\caption{Fundamental diagram for $\vm =2$ and $p=0.5$ (left) and $p=0.1$
(right). Comparison of paradisical MF (full line) with results from computer 
simulations ($\bullet$) and the naive MF approximation (dotted line).}
\label{fig_v2}
\end{figure}

We have also calculated the pMF equations which in addition take into
account the second order GoE state (\ref{goe3}). One has to modify the
equations for $c_0$ and $c_1$ only (and the normalization) by replacing
the last term in (\ref{c0eq}) by $pd\{(c_0+c_1)c+c_0d(1-c_2)\}$
and the last term in (\ref{c1eq}) by $qd\{(c_0+c_1)c+c_0d(1-c_2)\}$.
However, these modifications lead only to minor changes in the results
for the fundamental diagram.

\section{Discussion}

We have presented an analysis of the allowed configurations in the
CA model for traffic flow. Due to the use of parallel dynamics not
all configurations can be reached through the dynamics. Eliminating
these Garden of Eden states allowed us to improve the results of
the naive mean-field theory considerably.

GoE states can be characterized locally. We identified all elementary
GoE configurations for $\vm=1$ and $\vm=2$. It turns out that for
$\vm=1$ it is sufficient to investigate only configurations of all
clusters of two cells. For $\vm=2$ the largest elementary GoE configuration
consists of four neighbouring cells. 

For $\vm =1$ the paradisical mean-field theory is able to reproduce
the exact solution. This implies that in the subspace without GoE 
states all configurations are equally probable. This has to be 
compared with random-sequential dynamics. Here {\em all} configurations
are equally probable and naive mean-field theory is exact.
This means that the strong short-ranged correlations found for
parallel update are solely due to the use of parallel dynamics.

In fact one may speculate that this is rather general. The difference 
between random-sequential and parallel update comes mainly from the 
existence of GoE states in the latter. This implies that a method that 
'works' for random-sequential dynamics (e.g.\ an exact solution or good
approximation) should also work for parallel dynamics, but now in the
subspace without GoE states.

For $\vm =2$ the paradisical mean-field theory yields a considerable
improvement of the mean-field results, but it does not become exact.
One observes a qualitative difference to the case $\vm =1$, since
now there are correlations present which can not be explained by the 
existence of GoE states.

The existence of GoE states gives a simple criterion for the quality of
an approximation: A good approximation should be able to account for all 
GoE states. This can be illustrated for the case $\vm =1$. The methods
used previously for the exact solution are the 2-cluster approach 
\cite{ss,ssni} and Car-Oriented Mean-Field theory (COMF) \cite{comf}. Both 
methods are able to identify both GoE states (0,1) and (1,1). For $\vm=2$ 
one needs at least the 4-cluster approximation to account for all 
GoE states (\ref{goe1}-\ref{goe3}). Indeed, the results of \cite{ss,ssni} 
show that the 4-cluster results are in excellent agreement with simulation 
results.
COMF is able to identify all GoE states since all elementary GoE
configurations consist of only two neighbouring cars, i.e.\ there are no
elementary GoE states with three vehicles.

Finally we want to point out that the model is ergodic in the sense that 
for configurations $\tau$, $\tau'$ which appear in the stationary state 
with probabilities $P(\tau),P(\tau')>0$ there is a non-vanishing 
transition probability $P(\tau\to\tau')$. The existence of GoE states 
poses no problems in computer simulations. If the initial state
is a GoE state, it will become a non-GoE state after the first timestep.

The method presented here is also applicable to other models. An interesting
case is the asymmetric exclusion process (see \cite{rsss} and references
therein), which is identically to the model investigated here with $\vm=1$, 
but with open boundary conditions where an injection and/or removal of 
particles is possible. In \cite{ernst} the existence of GoE states has 
been used to obtain an approximative description of the deterministic limit
of the ASEP which is in excellent agreement with numerical results.


\noindent
{\bf Acknowledgements:} Part of this work has been performed within
the research program of the Sonderforschungsbereich 341 
(K\"oln-Aachen-J\"ulich). 


\end{document}